# SEEDS OF LIFE IN SPACE (SOLIS). III. ZOOMING INTO THE METHANOL PEAK OF THE PRE-STELLAR CORE L1544[*]

Anna Punanova,[1,2] Paola Caselli,[1] Siyi Feng,[1] Ana Chacón-Tanarro,[1] Cecilia Ceccarelli,[3,4] Roberto Neri,[5] Francesco Fontani,[6] Izaskun Jiménez-Serra,[7] Charlotte Vastel,[8,9] Luca Bizzocchi,[1] Andy Pon,[10] Anton I. Vasyunin,[1,2] Silvia Spezzano,[1] Pierre Hily-Blant,[3,4] Leonardo Testi,[6,11] Serena Viti,[12] Satoshi Yamamoto,[13,14] Felipe Alves,[1] Rafael Bachiller,[15] Nadia Balucani,[16] Eleonora Bianchi,[6,17] Sandrine Bottinelli,[8,9] Emmanuel Caux,[8,9] Rumpa Choudhury,[1] Claudio Codella,[6] François Dulieu,[18] Cécile Favre,[6] Jonathan Holdship,[12] Ali Jaber Al-Edhari,[3,4,19] Claudine Kahane,[3,4] Jake Laas,[1] Bertrand LeFloch,[3,4] Ana López-Sepulcre,[3,5] Juan Ospina-Zamudio,[3] Yoko Oya,[13] Jaime E. Pineda,[1] Linda Podio,[6] Davide Quenard,[7] Albert Rimola,[20] Nami Sakai,[21] Ian R. Sims,[22] Vianney Taquet,[23] Patrice Theulé,[24] and Piero Ugliengo[25]

[1]*Max-Planck-Institut für extraterrestrische Physik, Giessenbachstrasse 1, 85748 Garching, Germany*
[2]*Ural Federal University, 620002, 19 Mira street, Yekaterinburg, Russia*
[3]*IPAG, Université Grenoble Alpes, F-38000 Grenoble, France*
[4]*CNRS, IPAG, F-38000 Grenoble, France*
[5]*Institut de Radioastronomie Millimétrique, 300 rue de la Piscine, 38406, Saint-Martin dHéres, France*
[6]*INAF-Osservatorio Astrofisico di Arcetri, Largo E. Fermi 5, I-50125, Florence, Italy*
[7]*School of Physics and Astronomy, Queen Mary University of London, 327 Mile End Road, London, E1 4NS, UK*
[8]*Université de Toulouse, UPS-OMP, IRAP, Toulouse, France*
[9]*CNRS, IRAP, 9 Av. Colonel Roche, BP 44346, F-31028 Toulouse Cedex 4, France*
[10]*Department of Physics and Astronomy, The University of Western Ontario, 1151 Richmond Street, London, N6A 3K7, Canada*
[11]*European Southern Observatory, Karl-Schwarzschild-Str. 2, 85748 Garching bei München, Germany*
[12]*Department of Physics and Astronomy, University College London, Gower St., London, WC1E 6BT, UK*
[13]*Department of Physics, The University of Tokyo, Bunkyo-ku, Tokyo 113-0033, Japan*
[14]*Research Center for the Early Universe, The University of Tokyo, 7-3-1, Hongo, Bunkyo-ku, Tokyo 113-0033, Japan*
[15]*Observatorio Astronómico Nacional (OAN, IGN), Calle Alfonso XII, 3, 28014 Madrid, Spain*
[16]*Dipartimento di Chimica, Biologia e Biotecnologie, Università di Perugia, Via Elce di Sotto 8, I-06123 Perugia, Italy*
[17]*Dipartimento di Fisica e Astronomia, Università degli Studi di Firenze, Italy*
[18]*LERMA, Université de Cergy Pontoise, Sorbonne Universités, UPMC Univ. Paris 6, PSL Research University, Observatoire de Paris, UMR 8112 CNRS, 95000 Cergy Pontoise, France*
[19]*University of AL-Muthanna, College of Science, Physics Department, AL-Muthanna, Iraq*
[20]*Departament de Química, Universitat Autònoma de Barcelona, E-08193 Bellaterra, Spain*
[21]*The Institute of Physical and Chemical Research (RIKEN), 2-1, Hirosawa, Wako-shi, Saitama 351-0198, Japan*
[22]*Institut de Physique de Rennes, UMR CNRS 6251, Université de Rennes 1, 263 Avenue du Général Leclerc, F-35042 Rennes Cedex, France*
[23]*Leiden Observatory, Leiden University, P.O. Box 9513, 2300-RA Leiden, The Netherlands*
[24]*Aix-Marseille Université, PIIM UMR-CNRS 7345, 13397 Marseille, France*
[25]*Dipartimento di Chimica and NIS Centre, Università degli Studi di Torino, Via P. Giuria 7, I-10125 Torino, Italy*

## ABSTRACT

Corresponding author: Anna Punanova
punanova@mpe.mpg.de

[*] This work is based on observations carried out under project number L15AA with the IRAM NOEMA Interferometer and on observations carried out with the IRAM 30 m telescope. IRAM is supported by INSU/CNRS (France), MPG (Germany) and IGN (Spain).

Towards the pre-stellar core L1544, the methanol (CH$_3$OH) emission forms an asymmetric ring around the core centre, where CH$_3$OH is mostly in solid form, with a clear peak 4000 au to the north-east of the dust continuum peak. As part of the NOEMA Large Project SOLIS (Seeds of Life in Space), the CH$_3$OH peak has been spatially resolved to study its kinematics and physical structure and to investigate the cause behind the local enhancement. We find that methanol emission is distributed in a ridge parallel to the main axis of the dense core. The centroid velocity increases by about 0.2 km s$^{-1}$ and the velocity dispersion increases from subsonic to transonic towards the central zone of the core, where the velocity field also shows complex structure. This could be indication of gentle accretion of material onto the core or interaction of two filaments, producing a slow shock. We measure the rotational temperature and show that methanol is in local thermodynamic equilibrium (LTE) only close to the dust peak, where it is significantly depleted. The CH$_3$OH column density, $N_{tot}$(CH$_3$OH), profile has been derived with non-LTE radiative transfer modelling and compared with chemical models of a static core. The measured $N_{tot}$(CH$_3$OH) profile is consistent with model predictions, but the total column densities are one order of magnitude lower than those predicted by models, suggesting that the efficiency of reactive desorption or atomic hydrogen tunnelling adopted in the model may be overestimated; or that an evolutionary model is needed to better reproduce methanol abundance.

*Keywords:* Stars: formation – ISM: kinematics and dynamics – ISM: clouds – ISM: molecules – ISM: individual objects: L1544 – Radio lines: ISM

# 1. INTRODUCTION

Methanol ($CH_3OH$) is a crucial molecule for the growth of molecular complexity in the interstellar medium, as it is a key precursor for many organic and pre-biotic molecules found in regions of star- and planet formation (e.g. Herbst & van Dishoeck 2009). Methanol is widespread in our Galaxy and it is present in different environments, such as the molecular envelopes surrounding low-mass and high-mass protostars (the so-called hot corinos and hot cores), and cold dense clouds in low- and high- mass star-forming regions (e.g. Blake et al. 1987; Gibb et al. 2000; Schöier et al. 2002; Parise et al. 2004; Maret et al. 2005; Tafalla et al. 2006; Bizzocchi et al. 2014; Vastel et al. 2014). According to present models and experiments, methanol is formed on dust grains via the hydrogenation of CO (e.g. Tielens & Hagen 1982; Watanabe & Kouchi 2002; Rimola et al. 2014) and released to the gas phase via thermal and/or non-thermal processes (Garrod & Herbst 2006; Vasyunin & Herbst 2013). In the cold ($\simeq$10 K) dense ($10^4$–$10^7$ cm$^{-3}$) gas of pre-stellar cores, thermal desorption is not effective and reactive desorption is thought to be responsible for the release of methanol into the gas phase upon formation on icy mantles (e.g. Garrod et al. 2007; Vasyunin & Herbst 2013), in particular on CO-rich surfaces (Minissale et al. 2016b; Vasyunin et al. 2017). The photo desorption of methanol is not effective as it breaks the molecule into fragments, as has been shown experimentally (Cruz-Diaz et al. 2016; Bertin et al. 2016).

In dense cold cores, gaseous methanol should preferentially be found in a shell around the dense central regions, where visual extinctions are large enough to screen interstellar UV photons ($\geq$10 mag) and volume densities are around a few $\times 10^4$ cm$^{-3}$ (Vasyunin et al. 2017). In these conditions, carbon atoms are mainly locked in CO molecules and CO freeze-out becomes significant (Caselli et al. 1999; Tafalla et al. 2002). Methanol is then produced via surface hydrogenation of the frozen CO molecules and it is partially returned to the gas phase upon formation on CO-rich ices (Vasyunin et al. 2017). At higher densities, i.e. towards the core centre, the freeze-out rate of methanol overcomes its production rate, with a consequent drop in its gas-phase abundance (Vasyunin et al. 2017). Thus, gas-phase methanol is expected to be abundant at the edge of the CO-depleted zone. In fact, observations of methanol towards dense cores (L1498, L1517B; Tafalla et al. 2006) reveal ring-like structures.

In this work, we focus on the methanol emission towards L1544. This is a prototypical pre-stellar core, being centrally concentrated (Ward-Thompson et al. 1999), with a central density of $2\times10^6$ cm$^{-3}$, low central temperatures ranging from 5 to 11 K in the inner 10000 au (Crapsi et al. 2007) and undergoing a slow quasi-static contraction (Tafalla et al. 1998; Keto & Caselli 2010; Keto et al. 2015). It presents the chemical features of CO freeze-out and enhanced deuteration towards the centre (Caselli et al. 1999, 2002b; Vastel et al. 2006). L1544 also shows signs of chemical differentiation, with methanol residing away from the sharp $H_2$ column density drop towards the south-east of the core, rich in carbon chain molecules (Spezzano et al. 2016, 2017). $CH_3OH$ towards L1544 has also been found to have an asymmetric ring-like distribution, with the peak located towards the north-east of the dust peak (see Fig. 1), away from the low extinction regions (Bizzocchi et al. 2014; Spezzano et al. 2016). Several complex organic molecules (e.g. acetaldehyde, formic acid, dimethyl ether, methyl formate) have been detected towards L1544 (Vastel et al. 2014; Jiménez-Serra et al. 2016). At the location of the methanol peak of L1544, Jiménez-Serra et al. (2016) found enhanced abundances of O-bearing complex organic molecules (in particular $CH_3CHO$, $HCOOCH_3$, and $CH_3OCH_3$), likely related to methanol (also HCO, Spezzano et al. 2017), as well as $CH_3O$, a possible product of methanol photodissociation (Bertin et al. 2016; Cruz-Diaz et al. 2016) or, alternatively, a product of rapid gas-phase reactions between methanol and hydroxyl radical (OH) (Shannon et al. 2014).

We present interferometric observations of the methanol peak of L1544, with the aim of investigating its origins. This work is part of the NOEMA (Northern Extended Millimetre Array) large program SOLIS (Seeds of Life in Space), aimed at studying the formation of complex organic molecules across all stages of star formation (Ceccarelli et al. 2017). In section 2, the details of the observations, the data reduction procedure and Gaussian fitting of the spectra are presented. Section 3 presents the results of the Gaussian fitting, velocity gradients, rotational temperatures and column density calculations. In section 4, we discuss the results and possible origins of the methanol-rich zone. The summary of the paper is given in section 5.

# 2. OBSERVATIONS, DATA REDUCTION AND LINE FITTING

## 2.1. Observations

Observations of the $(2_{1,2}-1_{1,1})$-$E_2$, $(2_{0,2}-1_{0,1})$-$A^+$, and $(2_{0,2}-1_{0,1})$-$E_1$ methanol lines at $\simeq$96.74 GHz towards the methanol emission peak near L1544 (J2000 $\alpha = 05^h04^m18^s.0$, $\delta = +25°11'10''$, Bizzocchi et al. 2014) were carried out with the NOEMA interferometer in C and D configurations on 21–23 and 30 July and 25–26 October 2015 under average weather conditions (pwv=1–10 mm). The rest frequencies are given in Table 1. The primary beam size was 52″, the synthesized beam was 5.71×3.86″ at a position angle $\theta = -52.48°$. The data were obtained with the narrow band correlator with a spectral resolution of 39 kHz, corresponding to a velocity resolution of 0.12 km s$^{-1}$. The system temperatures were 70–250 K. Sources 0234+285, MWC349, LKHA101, and 0507+179 were used as flux calibrators; 0507+179 was used as a phase/amplitude calibrator; and 3C454.3 and 3C84 were used as bandpass calibrators.

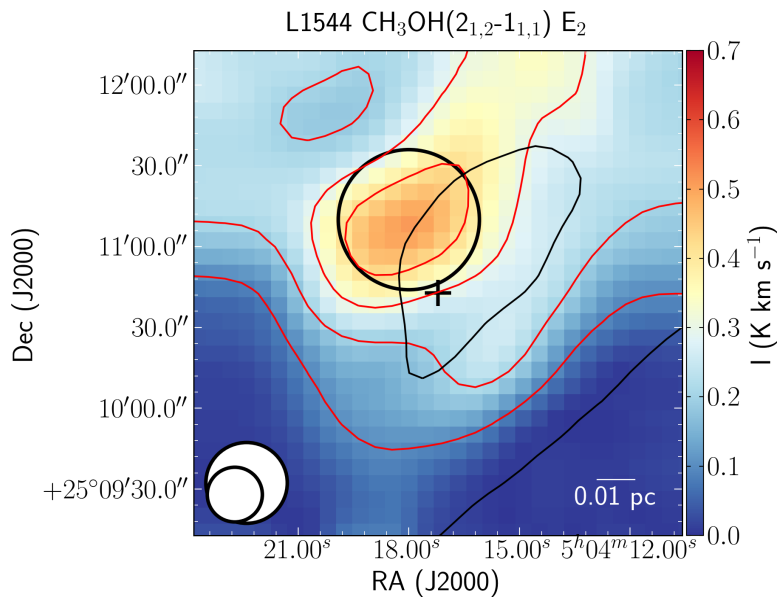

**Figure 1.** Methanol emission towards L1544 mapped with the IRAM 30 m antenna (color scale, Bizzocchi et al. 2014) and the 250 μm dust continuum emission mapped with *Herschel*/SPIRE (black contours, André et al. 2010). The red contours of methanol start at 10 $\sigma_I$ (0.1 K km s$^{-1}$) with a step of 10 $\sigma_I$. The thin black contours show the 50% and 90% of the peak dust emission (232.81 MJy sr$^{-1}$). The thick black circle shows the NOEMA primary beam. The *Herschel* and the 30 m beams are shown in the bottom left (the larger beam is of *Herschel*, the smaller beam is of the 30 m). The 1.3 mm dust continuum emission peak (Ward-Thompson et al. 1999), considered as the core centre, is shown with the cross.

Simultaneously, the dimethyl ether (CH$_3$OCH$_3$ ($5_{5,1}$–$4_{4,0}$)-$EA$ at 95.85 GHz) and methyl formate (CH$_3$OCHO ($5_{4,1}$–$5_{3,3}$)-$E$ at 96.94 GHz and ($17_{5,12}$–$17_{4,13}$)-$A$ at 97.20 GHz) lines were observed with the same spectral setup with the narrow band correlator but were not detected (the rms was 6 mJy beam$^{-1}$ with the synthesized beam size of 5.7″×3.9″). Dust continuum emission observed with the wide band correlator WideX was not detected down to an rms noise level of 0.026 mJy beam$^{-1}$ with a beam size of 3.4″×2.4″. The spectral range of WideX was 95.85–99.45 GHz. The SO($2_3$–$1_2$) line at 99.30 GHz and the CS(2–1) line at 97.98 GHz lines were detected in the WideX band with a spectral resolution of 1950 kHz or 5.9 and 6.0 km s$^{-1}$ at the given frequencies, respectively. The peak intensities were ∼4.0 mJy beam$^{-1}$ and ∼3.5 mJy beam$^{-1}$ for CS and SO, respectively, and the rms was 0.7 mJy beam$^{-1}$ with a beam size of 3.4″×2.4″[1]. Because of their poor spectral resolution compared to the methanol lines, these data will not be discussed in this paper.

To recover the emission from scales larger than 20″ we combined the synthetic visibilities derived from the IRAM 30 m observations of the methanol lines obtained by Bizzocchi et al. (2014) with our NOEMA data. The single dish observations were carried out in October 2013 under excellent weather conditions (pwv≃0.5 mm). The on-the-fly maps were obtained with the EMIR 090 (3 mm band) heterodyne receiver in position switching mode, using the FTS backend with a spectral resolution of 50 kHz; this corresponds to a velocity resolution of 0.15 km s$^{-1}$ at the frequency of 96.74 GHz. The angular resolution was 25.6″. The 3′ × 3′ maps were centred at the dust emission peak (J2000 $\alpha = 05^h04^m17^s.21$, $\delta = +25°10′42.8″$). The pointing accuracy of the 30 m antenna was better than 1″. The system temperature was ≃90 K (for details see Bizzocchi et al. 2014).

### 2.2. *Data reduction: spectral data cubes*

---

[1] The sizes of the synthesized beams for methanol and other lines observed with the narrow band correlator and for continuum and the lines observed with WideX are different because for the narrow band correlator observations 6 antennas are used while for WideX observations 7–8 antennas are used. The number of antennas used impacts the uv coverage and the synthesized beam size.

Table 1. The observed methanol lines.

| Transition | Frequency[a] (GHz) | $E_{up}/k$ [a] (K) | $A$ [a] ($10^{-5}$ s$^{-1}$) | $n_{crit}^{(c)}$ ($10^5$ cm$^{-3}$) |
|---|---|---|---|---|
| $(2_{1,2}-1_{1,1})$-$E_2$ | 96.739362 | 12.53[b] | 0.2558 | 0.82 |
| $(2_{0,2}-1_{0,1})$-$A^+$ | 96.741375 | 6.96 | 0.3408 | 1.09 |
| $(2_{0,2}-1_{0,1})$-$E_1$ | 96.744550 | 20.08[b] | 0.3407 | 1.09 |

Notes. [a] The frequencies, energies and Einstein coefficients are taken from Bizzocchi et al. (2014) following Xu & Lovas (1997) and Lees & Baker (1968), also available at the JPL database (Pickett et al. 1998). [b] Energy relative to the ground $0_{0,0}$, A rotational state. [c] The critical densities are calculated for kinetic temperature of 10 K.

The calibration, imaging, and cleaning of the NOEMA data were performed with the CLIC and MAPPING packages of the GILDAS software[2]. The single dish data reduction up to the stage of convolved spectral data cubes was performed with the CLASS package of GILDAS. The comparison of the peak intensities of the methanol lines observed with the NOEMA and with the 30 m antenna shows that the interferometric observations recover 50–60% of the total flux. To recover the missing flux we merged the NOEMA and the 30 m data with a standard routine in the MAPPING package. The resulting data cubes have a velocity resolution of 0.15 km s$^{-1}$, the same as the single dish data. After the correction for the NOEMA primary beam response function, the rms of the resulting spectral data cubes varies from 0.003 in the centre to 0.009 Jy/beam at the edges of the primary beam. The synthesized beam of the combined data cube is 6.50″×4.06″ at a position angle $\theta = -49.95°$, with a pixel size 1.5″×1.5″.

### 2.3. Pyspeckit line fitting

The line fitting was performed with the Pyspeckit module of Python (Ginsburg & Mirocha 2011). The three methanol lines were fitted with a Gaussian profile in each pixel. The routine varies three parameters (peak intensity, centroid velocity $V_{LSR}$, and velocity dispersion $\sigma$) and finds the best fit with the Levenberg-Marquardt non-linear regression algorithm. The velocity dispersion was corrected for the channel width. The fit results were written to the final data cubes after masking poor data. In particular, for the integrated intensity maps, we used all data within the primary beam. For the centroid velocity and velocity dispersion we used the data within the primary beam, with a velocity dispersion accuracy better than 20% ($\sigma/\Delta\sigma > 5$), and with a high signal-to-noise ratio (SNR): $I > 5 \cdot rms \cdot \sqrt{N_{ch}} \cdot \Delta v_{res}$, where $I$ is the integrated intensity, $N_{ch}$ is the number of channels in the line, and $\Delta v_{res}$ is the velocity resolution. For $N_{ch}$ we take all channels in the range 6.1–8.0 km s$^{-1}$. This range defines the emission above one $rms \cdot \sqrt{N_{ch}} \cdot \Delta v_{res}$ over the spectrum averaged over the whole mapped area.

## 3. RESULTS

### 3.1. Distribution of methanol emission

Figure 2 shows the integrated intensity of the $E_2$ methanol line before combining the NOEMA data with the zero-spacing data from the 30 m antenna. Here, only the compact emission resolved by NOEMA is present. The methanol emission detected with NOEMA has an elongated structure on the northern side of the dense core with direction about perpendicular to the direction of the main axis of the dense core and to the structure seen in the NOEMA+30 m map (shown in Fig. 3); its thickness is about 10″ (1400 au at the distance of 140 pc) and the structure partly overlaps with the mm dust emission, from the 3 $\sigma_{S_{1.2\ mm}}$ up to the 6 $\sigma_{S_{1.2\ mm}}$ contour. The substructure might be due to a local abundance variation; the substructure orientation resembles the direction of the northern filament in the large scale dust continuum emission map displayed in Fig. 13 and it could be tracing the "contact point" between the two filaments.

The compact emission disappears partially after combining NOEMA with 30 m data, as shown in Fig. 3 for the brightest methanol line $A^+$ (the integrated intensity maps of the $E_1$ and $E_2$ lines are shown in the appendix; see Fig. A.1). This maps clearly shows a ridge elongated in approximately the same direction as the main axis of the dense core, with thickness of ∼3000 au; the ridge contains a well defined peak at $\alpha = 05^h04^m17.68^s$, $\delta = +25°11'08.3''$, with size ∼1300 au, and two secondary peaks south-east of the main peak. This ridge could be the zone where the two filaments are interacting and/or where cloud material is accreting onto the dense core. The map is centred at the methanol emission peak revealed with the IRAM 30 m observations (Bizzocchi et al. 2014). The dust emission peak at 1.3 mm from Ward-Thompson et al. (1999) (considered as the core centre) is shown with a black cross in Fig. 3, outside the primary beam area. Chemical models of spherically symmetric pre-stellar cores predict that gas-phase methanol should be found in a shell around the dense core central regions (e.g. Vasyunin et al. 2017), giving rise to a ring-like structure in observations (e.g. Tafalla et al. 2006). Although the methanol emission is distributed around the dust continuum peak of L1544 (see Bizzocchi et al. 2014, and Fig. 1), the ring-like structure is not uniform, with a clear maximum about 4000 au to the north-east of the dust peak position. The asymmetric distribution of methanol could be related to the inhomogeneities in the distribution of cloud material around the dense core, with the southern part more exposed to the interstellar radiation field (Spezzano et al. 2016). Gas phase methanol preferentially traces the more shielded material around the dense core, where carbon is mainly locked in CO molecules.

---

[2] The GILDAS software is developed at the IRAM and the Observatoire de Grenoble, and is available at http://www.iram.fr/IRAMFR/GILDAS

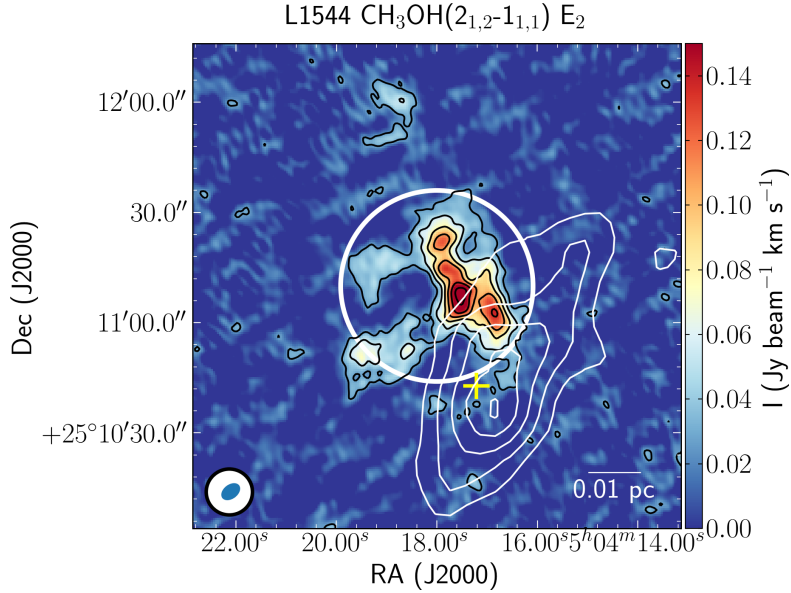

**Figure 2.** Integrated intensity of the $E_2$ line before combining with the single dish data (the map is not corrected for the primary beam attenuation). The black contours start at 3 $\sigma_S$ (0.027 Jy beam$^{-1}$) with a step of 3 $\sigma_S$. The white contours represent the 1.2 mm dust continuum emission from NIKA (Chacón-Tanarro et al. 2017). They start at 3 $\sigma_S$ (0.021 Jy beam$^{-1}$, with the beam size of 12.5″) and increase with a step of 1 $\sigma_S$. The white circle in the centre is the primary beam of NOEMA for the methanol data. The yellow cross shows the dust emission peak (Ward-Thompson et al. 1999). The synthesized beam of NOEMA (blue) and the NIKA beam (white) are shown in the bottom left corner.

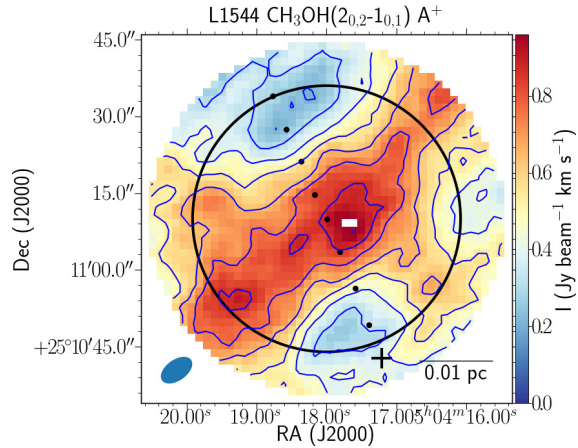

**Figure 3.** Integrated intensity of the $A^+$ methanol line (NOEMA+30 m). The blue contours represent integrated intensity, and start at 0.216 Jy beam$^{-1}$ km s$^{-1}$ with a step of 0.108 Jy beam$^{-1}$ km s$^{-1}$. $3\sigma_I$=0.005 Jy beam$^{-1}$ km s$^{-1}$. The white circle is the NOEMA primary beam. The black cross shows the dust emission peak (Ward-Thompson et al. 1999). The synthesized beam of NOEMA is shown in the bottom left corner. The black dots show the positions of spectra used for non-LTE modelling (see Sect. 3.5 for details). The white pixels are those masked because of low quality spectra.

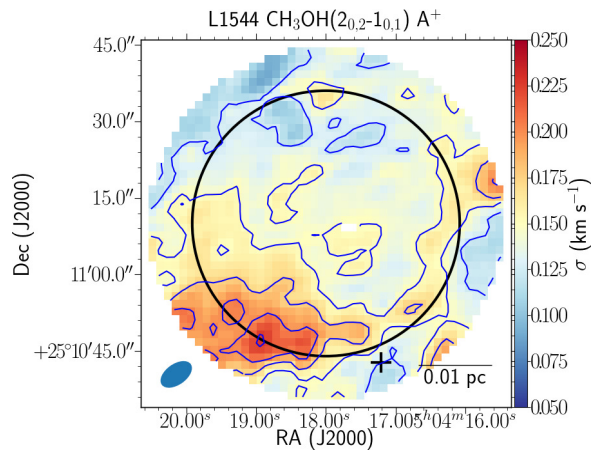

**Figure 4.** Velocity dispersions of the $A^+$ methanol line. The blue contours show velocity dispersions of 0.125, 0.150, 0.175, 0.200, and 0.225 km s$^{-1}$. The black circle shows the primary beam. The cross shows the 1.3 mm dust emission peak (Ward-Thompson et al. 1999). The synthesized beam is shown in the bottom left corner. The white pixels are those masked because of low quality spectra.

### 3.2. *Kinematics*

#### 3.2.1. *Velocity dispersion*

Figures 4 and A.2 show the velocity dispersions ($\sigma = \Delta v/\sqrt{8\ln(2)}$, where $\Delta v$ is the full width at half maximum, FWHM) of the methanol lines. The velocity dispersions range from 0.11 to 0.26 km s$^{-1}$ with a median value of 0.15 km s$^{-1}$ and typical uncertainties of 0.008 km s$^{-1}$ and 0.014 km s$^{-1}$ for the bright lines ($A^+$ and $E_2$) and weak line ($E_1$). The line width increases towards the south-east clearly in the $A^+$ and $E_2$ lines, and tentatively also in the $E_1$ line.

Towards the south-east, the velocity dispersions in the (NOEMA+30 m) combined map are larger than those observed with the 30 m single-dish telescope by 0.05–0.10 km s$^{-1}$. This difference is not systematic: it decreases with distance from the location with the largest dispersion in the south-east of the core and becomes negligible in the northern part of the map. A detailed inspection of the spectra reveals the presence of a small-scale higher velocity part of the line in the south-eastern part of the core. NOEMA, being more sensitive than the 30 m antenna, reveals a weak higher velocity component, associated with the small scale structure seen by NOEMA-only (see Fig. 2), too small to be detected with the large beam of the single dish. In the combined spectrum, the resulting line is slightly broader and its centroid velocity is slightly larger than that observed with the single dish. The high-velocity part of the line detected with NOEMA appears towards the southern part of the primary beam, closer to the densest regions of the pre-stellar core.

#### 3.2.2. *Non-thermal motions*

Figure 5 shows the ratio of the non-thermal components ($\sigma_{NT}$) of the three methanol lines in each pixel within the primary beam and the thermal velocity dispersion of a mean particle, $\sigma_T$, as a function of the distance to the dust peak. The non-thermal components are derived from the observed velocity dispersion ($\sigma_{obs}$) via

$$\sigma_{NT}^2 = \sigma_{obs}^2 - \frac{kT_k}{m_{obs}}, \tag{1}$$

where $k$ is Boltzmann's constant, $T_k$ is the kinetic temperature, and $m_{obs}$ is the mass of the observed molecule. The formula is adopted from Myers et al. (1991). The thermal velocity dispersion of a mean particle is $\sigma_T = \sqrt{kT_k/\mu}$, where $\mu = 2.37$ amu is the mean particle mass (Kauffmann et al. 2008). We assume that the kinetic temperature is 10 K as this is the temperature measured with ammonia by Crapsi et al. (2007) at the distance of the methanol peak. This temperature is also consistent with the methanol rotational temperatures towards the high density gas close to the dust peak, where methanol is close to local thermodynamic equilibrium (LTE; see Sect. 3.3 for details).

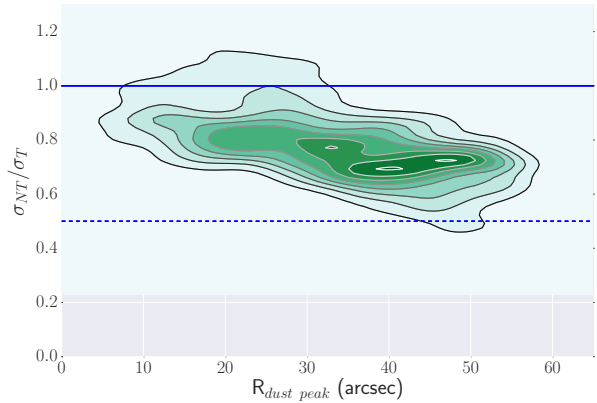

**Figure 5.** Ratio of non-thermal components of the three methanol lines to the thermal line width of a mean particle as a function of distance from the dust peak. The solid and dashed blue horizontal lines show the $\sigma_{NT}/\sigma_T$ ratios equal to 1 and 0.5. The colour scale and grey contours represent the number of the data points. Only the data points within the primary beam are used for the plot. The grey area does not contain any data points.

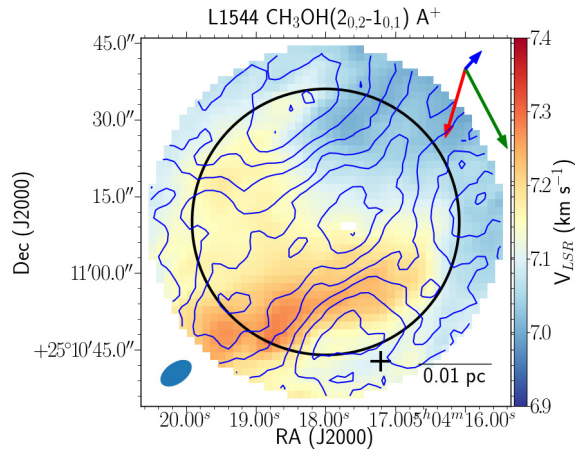

**Figure 6.** Centroid velocities of the $A^+$ methanol line. The blue contours represent the integrated intensity, starting at 0.216 Jy beam$^{-1}$ km s$^{-1}$ with a step of 0.108 Jy beam$^{-1}$ km s$^{-1}$; $3\sigma_I$=0.005 Jy beam$^{-1}$ km s$^{-1}$. The red arrow shows the total velocity gradient measured with the methanol line ($7.10\pm0.01$ km s$^{-1}$ pc$^{-1}$). The green and blue arrows show the total velocity gradients of NH$_3$ and NH$_2$D measured at scales of 15000 au and 4000 au, respectively (Crapsi et al. 2007) (the length of the arrow is proportional to the velocity gradient magnitude). The circle shows the primary beam. The black cross shows the 1.3 mm dust emission peak (Ward-Thompson et al. 1999). The synthesized beam is shown in the bottom left corner. The white pixels are those masked because of low quality spectra.

The thermal velocity dispersion ($\sigma_T$) for a mean particle with mass 2.37 amu at 10 K is 0.19 km s$^{-1}$. The ratio of the non-thermal component to the thermal velocity dispersion varies from 0.3 to 1.7, being 0.8 on average. The ratio decreases with distance from the dust peak from $\sim$0.9 to $\sim$0.7 and it reaches unity towards the south-east. The majority of the lines (92%) are subsonic, with the small fraction of transonic lines coming from the south-east region.

### 3.2.3. *Velocity field*

Figure 6 and A.3 show the centroid velocity ($V_{LSR}$) maps. The $V_{LSR}$ varies in the range 6.9–7.3 km s$^{-1}$. The interferometric observations reveal substructure in the velocity field, with the velocity increasing towards the south, south-east and east. The three lines show similar velocity patterns; the $E_1$ line shows higher velocities in the southeast, where the SNR of the weak $E_1$ line is low ($\simeq 5$) compared to the $E_2$ and $A^+$ lines which SNR there are $> 20$. We estimate total and local velocity gradients across the methanol emission following the method described in Goodman et al. (1993) for total gradients and applied for local gradients by Caselli et al. (2002a) (see the description of local gradients below). The total velocity gradient calculation provides the average velocity across the mapped region, $< V_{LSR} >$, the magnitude of the velocity gradient, $G$, and the position angle, $\theta_G$. The total gradients are calculated using all available points weighted by $1/\Delta^2_{V_{LSR}}$, where $\Delta_{V_{LSR}}$ is the uncertainty of the centroid velocity.

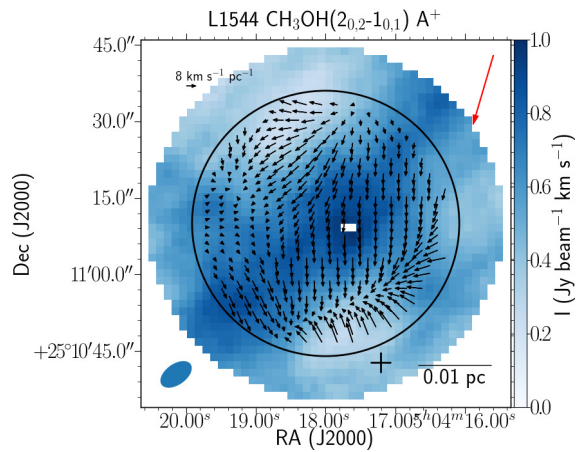

**Figure 7.** Local velocity gradients of the $A^+$ methanol line. The color scale shows integrated intensity. The black arrows indicate the local velocity gradients. The red arrow shows the total velocity gradient (the scale of the total gradient is eight times larger than the scale of the local gradients). The circle shows the primary beam. The cross denotes the 1.3 mm dust emission peak (Ward-Thompson et al. 1999). The synthesized beam is plotted in the bottom left corner. The white pixels are those masked because of low quality spectra.

The total gradients have been measured for the three methanol lines: $G = 8.77 \pm 0.04$, $7.10 \pm 0.01$, and $7.24 \pm 0.01$ km s$^{-1}$ pc$^{-1}$, with $\theta_G = 153.7° \pm 0.3°$, $162.13° \pm 0.08°$, and $163.76° \pm 0.07°$ measured east of north, for the $E_1$, $A^+$, and $E_2$ lines, respectively. The total gradient for the $A^+$ line is shown as red arrows in Fig. 6 and 7. For comparison, the green and blue arrows in Fig. 6 represent the velocity gradients measured with the high density tracers NH$_3$ and NH$_2$D using interferometric data across the entire L1544 core (Crapsi et al. 2007) (with the arrow size proportional to the magnitude of the velocity gradient). The NH$_3$ traces the whole dense core (radius $\sim 15000$ au), while NH$_2$D traces the inner region of the core, which includes the dust peak ($\sim 4000$ au; see Crapsi et al. 2007, for details). Williams et al. (2006) also measure the total gradient of the dense core with the interferometric observations of N$_2$H$^+$ ($G = 4.1$ km s$^{-1}$ pc$^{-1}$). The gradient direction found using methanol significantly differs from those of the dense core tracers ($\theta_G$ differs by $\sim 20°$, $\sim 40°$, and $\sim 140°$ from those of N$_2$H$^+$, NH$_3$, and NH$_2$D, respectively) so we can conclude that the shell traced by methanol is not following the dense core kinematics. It is also interesting to compare the total velocity gradient deduced by methanol to the direction of the large-scale magnetic field measured in the north part of L1544 ($\theta_B = 30°$; Clemens et al. 2016). The minimum difference between the total velocity gradient direction at the scale of the NOEMA methanol map and the magnetic field direction is $\simeq 50°$, thus showing gas motions not aligned with larger scale magnetic field directions; this may indicate changes in the magnetic field direction towards the dense regions of the pre-stellar core as also pointed out by Clemens et al. (2016).

The local velocity gradients are presented in Fig. 7 and A.4 with black arrows, plotted over integrated intensity colour maps, along with red arrows which present the total velocity gradients. To calculate a gradient in a local position, we use all pixels within 6″ (where 1 pixel is 1.5″ in size), weighted according to their distance to the given position and their centroid velocity uncertainty:

$$w = \frac{1}{\Delta^2_{V_{LSR}}} \cdot \exp\left\{-d^2 / \left[2\left(\frac{\theta_{Gauss}}{2.354}\right)^2\right]\right\}, \quad (2)$$

where $w$ is the weight, $\Delta_{V_{LSR}}$ is the centroid velocity uncertainty, $d$ is the distance from the weighted pixel to the given position, and $\theta_{Gauss} = 4$ pixels is the FWHM of the weighting function. The four-pixel radius is used to compensate for the oversampling of the map. The typical errors for the local gradient values are 1 km s$^{-1}$ pc$^{-1}$ for the $E_1$ line and 0.2 km s$^{-1}$ pc$^{-1}$ for the $E_2$ and $A^+$ lines; and for the position angles of the local gradients – 4° for the $E_1$ line and 1.5° for the $E_2$ and $A^+$ lines.

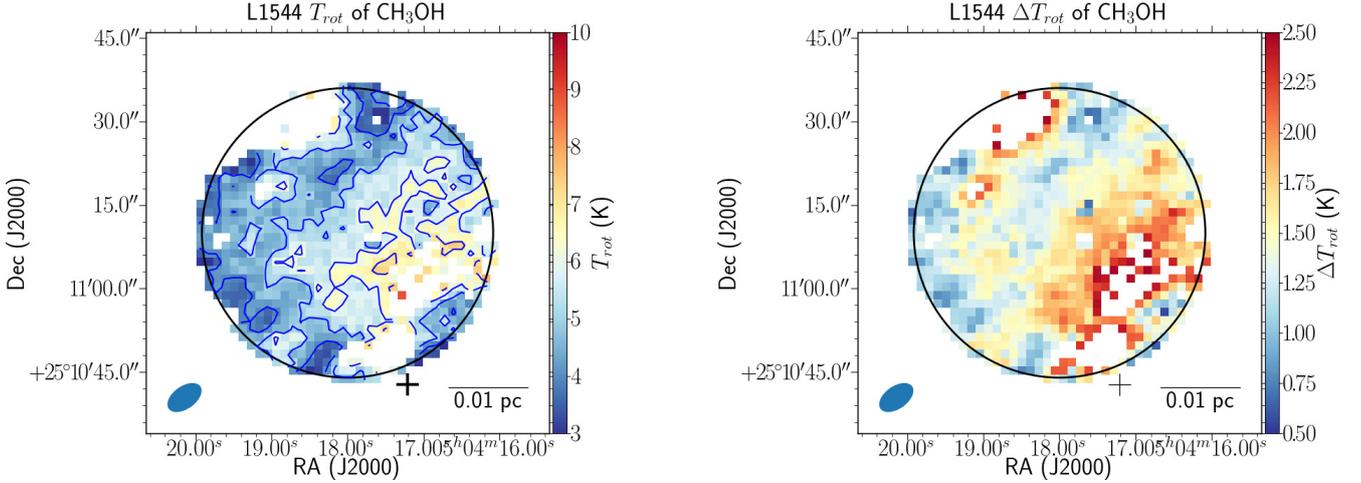

**Figure 8.** Rotational temperature of methanol (left) and its uncertainty (right). The blue contours on the left panel show $T_{rot}$ of 4, 5, 6, 7, and 8 K. The cross shows the dust peak position. The synthesized beam is shown in the bottom left corner. The white pixels are those where the $E_1$ line was not detected or those masked because of low $T_{rot}$ accuracy ($\Delta T_{rot} \geq 2.5$ K).

The velocity increases towards the south-east, as shown with the total velocity gradient direction. However, the arrows now indicate that the velocity field is not smooth and quite complex, with local velocity gradients showing significant variations in magnitude and direction across the observed area. The local velocity gradient values vary from ≃0.5 to 12 km s$^{-1}$ pc$^{-1}$. There is a high velocity bar in the south of the mapped area and then a sharp decrease (local velocity gradients are ≃11 km s$^{-1}$ pc$^{-1}$) of the centroid velocity further to the south, towards the dust peak (see Fig. 6 and 7). This sudden inversion of the direction in velocity gradient could be reproduced at the intersection of flows moving in different directions along the line of sight, which could support the scenario of the gentle collision between the two large-scale filaments clearly seen in the *Herschel* map shown in Fig. 13. However, one cannot exclude complex kinematics due to accretion of material at the north-eastern edge of the quiescent pre-stellar core.

### 3.3. *Rotational temperature*

Using the spectra of the observed lines we calculate the rotational temperature $T_{rot}$ and the total column density $N_{tot}$ of methanol, assuming LTE and optically thin emission. We assume $A : E$ methanol ratio to be 1 : 1. With the assumption of LTE, the population of all the energy levels can be described by a unique temperature, $T_{rot}$. With the assumption of optically thin emission, $T_{rot}$ is defined as $-1/a$ from a linear fit $ax + b$ to a $\log(N_{up}/g_{up})$ versus $E_{up}$ plot (rotational diagram), where $E_{up}$ is the energy of the upper level, expressed in K (given in Table 1); $N_{up}$ is the column density of the upper level population, defined as

$$N_{up} = \frac{8\pi k W \nu^2}{A h c^3}, \qquad (3)$$

where $k$ is the Boltzmann constant, $W$ is the integrated intensity of the line, $\nu$ is the frequency, $A$ is the Einstein coefficient (given in Table 1), $h$ is the Planck constant, and $c$ is the speed of light (e.g. Goldsmith & Langer 1999).

Figure 8 shows the map of rotational temperature (left) and its uncertainty (right). We show only those values with an uncertainty $\Delta T_{rot} < 2.5$ K. The typical values for $\Delta T_{rot}$ are 1–2 K. $T_{rot}$ varies from 3.0±0.8 K to 9±2 K, with an average $T_{rot}$ of 5.3±1.0 K. The temperature increases towards the south-west and the dust peak. The rotational temperature increase is most likely a result of the gas volume density increase towards the core centre, from a few $10^4$ cm$^{-3}$ in the north-eastern part of the observed area to a few × $10^5$ cm$^{-3}$ in the south-western part (see e.g. the model of Keto & Caselli 2010), as the methanol lines have critical densities of ≃ $10^5$ cm$^{-3}$. As the density increases towards the core centre, the energy level populations become closer to those expected in LTE and the rotational temperature approaches the kinetic temperature of the gas.

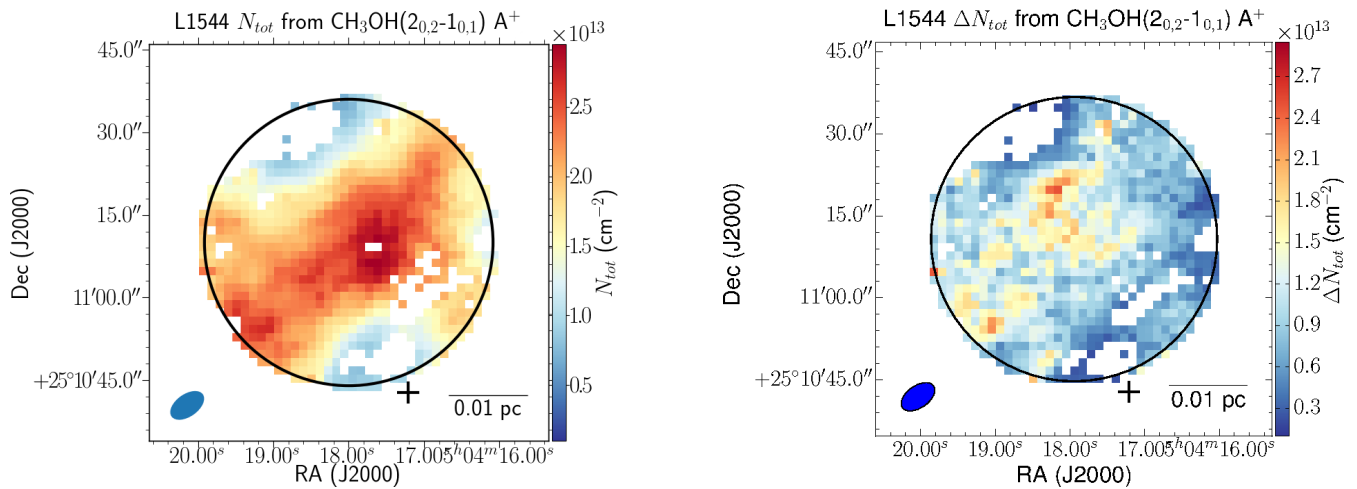

**Figure 9.** Total column densities of methanol measured with the $A^+$ line (left) and its uncertainty (right) derived with the assumption of LTE. The circle shows the primary beam. The cross shows the dust peak position. The synthesized beam is shown in the bottom left corner. The white pixels are those where the $E_1$ line was not detected or those masked because of low $T_{rot}$ accuracy ($\Delta T_{rot} \geq 2.5$ K).

Crapsi et al. (2007), derived ammonia rotational temperatures towards the core centre and found that the temperature increases from the centre outwards, from 5.5 to 10–13 K. The ammonia map obtained with the VLA covers the ammonia emission area of $75'' \times 36''$ centred at the dust peak. The peak methanol rotational temperature of $9\pm 2$ K is consistent with these ammonia derived temperatures, suggesting that the methanol is indeed close to being in LTE in the south-western part of the mapped region. It is also consistent with the kinetic temperature derived with non-LTE modelling for methanol towards the L1544 dust peak by Vastel et al. (2014) (7–15 K). This result confirms that $CH_3OH$ is mainly tracing a shell around the L1544 dust peak, as towards the centre, Crapsi et al. (2007) found temperatures of $\sim$6 K using $NH_3$, which does not appear to freeze-out (see also Caselli et al. 2017). The opposing directions of the methanol and ammonia temperature gradients further support the notion that the methanol gradient is only an excitation effect caused by changes in the gas density. In general, $CH_3OH$ and $NH_3$ trace different material, as $NH_3$ (as well as other N-bearing molecules) remains in the gas phase at significantly higher volume densities compared to C-bearing molecules (e.g. Caselli et al. 1999; Tafalla et al. 2002; Hily-Blant et al. 2010; Bizzocchi et al. 2014), but $CH_3OH$ and $NH_3$ overlap at densities between $10^4$ and $10^5$ cm$^{-3}$, where $CH_3OH$ maintains a detectable abundance in the gas phase.

### 3.4. Column density and methanol abundance

The total column density is given by:

$$N_{tot} = \frac{N_{up} Q_{rot}}{g \exp(-E_{up}/kT)}, \quad (4)$$

where $g$ is the statistical weight of the upper level ($g_J = 2J + 1$, with $J$ being the rotational quantum number), $E_{up}$ is the energy of the upper level, $k$ is the Boltzmann constant, $T_{rot}$ is used as the temperature $T$, and $Q_{rot}$ is the rotational partition function; its values for different temperatures are taken from the CDMS database (Müller et al. 2001).

Figure 9 shows the total column density map of methanol (left) and its uncertainty (right) derived for the brightest line, $A^+$. The variation between $N_{tot}$ found using the different lines is within a factor of 3: $N_{tot}(E_2)/N_{tot}(E_1) \simeq 2$, $N_{tot}(E_2)/N_{tot}(A^+) \simeq 3$. The uncertainties of the total column densities $\Delta N_{tot}$ are high: 30–300%, 20–100%, and 20–300% for the $E_1$, $A^+$, and $E_2$ lines, respectively, so the differences between the total column densities defined with the different lines are within the errors. The average column densities over the $30''$ beam close to the dust peak within the primary beam are $(3.0\pm 0.9)\times 10^{13}$ cm$^{-2}$, $(2.1\pm 0.6)\times 10^{13}$ cm$^{-2}$, and $(5.7\pm 1.6)\times 10^{13}$ cm$^{-2}$ for the $E_1$, $A^+$, and $E_2$ lines, respectively. This result is consistent with that of Bizzocchi et al. (2014), who found $(2.7\pm 0.6)\times 10^{13}$ cm$^{-2}$, and that of Vastel et al. (2014), who found $(2.6$–$3.8)\times 10^{13}$ cm$^{-2}$, towards the dust peak observed with the IRAM 30 m telescope (beams of $30''$ and $26''$, thus also partially including the area mapped with NOEMA).

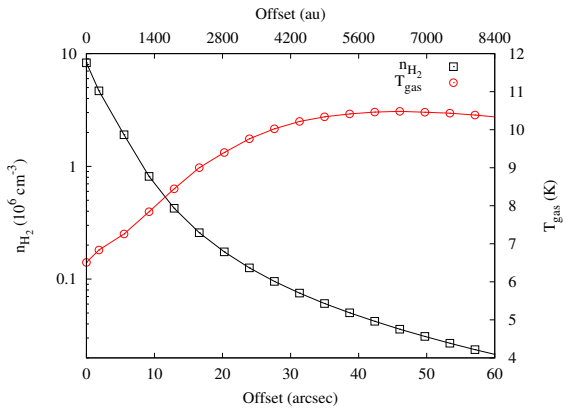

**Figure 10.** The physical structure of the modelled core: molecular hydrogen number density (black) and kinetic gas temperature (red).

We use the H$_2$ column density, $N$(H$_2$), to define the average methanol abundance within the primary beam area. The molecular hydrogen column density map for L1544 was produced by Spezzano et al. (2016) using the dust continuum emission data from the three *Herschel*/SPIRE bands at 250 $\mu$m, 350 $\mu$m, and 500 $\mu$m. As the *Herschel* beam of 38″ is comparable with the NOEMA primary beam at 96.4 GHz (52″) we can only define an average abundance of methanol within the observed area. With the average total column densities and molecular hydrogen column density (averaged over the 30″ beam close to the dust peak within the NOEMA primary beam area), $(2.3\pm0.3)\times10^{22}$ cm$^{-2}$, the average methanol abundances

$$X(\mathrm{CH_3OH}) = N_{tot}(\mathrm{CH_3OH})/N(\mathrm{H_2}) \qquad (5)$$

are $(1.3\pm0.4)\times10^{-9}$, $(0.9\pm0.3)\times10^{-9}$, and $(2.5\pm0.8)\times10^{-9}$ for the $E_1$, $A^+$, and $E_2$ lines, respectively, consistent with $0.92\times10^{-9}$ found by Bizzocchi et al. (2014) and lower than that found by Vastel et al. (2014), $6\times10^{-9}$. The difference between our result and that of Vastel et al. (2014) is due to the lower molecular hydrogen column density they assumed ($5\times10^{21}$ cm$^{-2}$).

### 3.5. *Non-LTE modelling*

To put better constraints on the column densities of methanol, taking into account the physical structure of the source, we perform non-LTE modelling of the methanol lines in the same manner as Bizzocchi et al. (2014). We use the radiative transfer code MOLLIE (Keto & Rybicki 2010) which produces line synthetic spectra based on the physical model of L1544 derived by Keto et al. (2014) (see Fig. 10 for the density and temperature profiles). L1544 is modelled as an unstable slowly contracting Bonnor-Ebert sphere with radiative heating and cooling of the gas and dust and simplified CO and H$_2$O chemistry to set the abundances of the major gas coolants (Keto & Caselli 2008; Keto et al. 2014). The CH$_3$OH abundance profile across the core is assumed to follow the abundance profile of the mother molecule, CO, with $X(\mathrm{CH_3OH}) = 0.35\times10^{-9}$ towards the dust peak. We chose eight positions on the line connecting the dust peak and the methanol peak within the primary beam, so the distance between the positions is similar to the synthesized beam size (the positions are shown with black dots in Fig. 3) to measure the methanol column densities along the core radius. Since methanol molecules do not switch their symmetries between the $A$ and $E$ forms, CH$_3$OH-$A$ and CH$_3$OH-$E$ may be considered as two different molecules with similar abundances. We compare independently the observed methanol $A$ and $E$ spectra with modelled lines and find the abundance profile which better reproduces the observed lines. The modelled spectra are presented in Fig. 11. The red lines represent the simulated spectra, after smoothing to the same velocity resolution as the observed spectra, which better resemble the observed spectra (black). The dark grey strips represent the modelled lines produced with one step smaller and one step larger column densities (one step is 10% of the given column density).

The resulting column density profiles (along the core radius through the methanol peak) are shown in Fig. 12. The column densities of CH$_3$OH-$A$ and CH$_3$OH-$E$ agree very well, within 10% which is consistent with the assumed 1:1 $E$:$A$ abundance ratio. The modelled column density varies between $\simeq0.7\times10^{13}$ cm$^{-2}$ close to the dust peak and $\simeq2.4\times10^{13}$ cm$^{-2}$ towards the methanol peak for each $A$- and $E$-methanol, that is the total column density of methanol is $(1.4–4.8)\times10^{13}$ cm$^{-2}$. This result agrees very well with that calculated assuming LTE (see Sect. 3.4 and

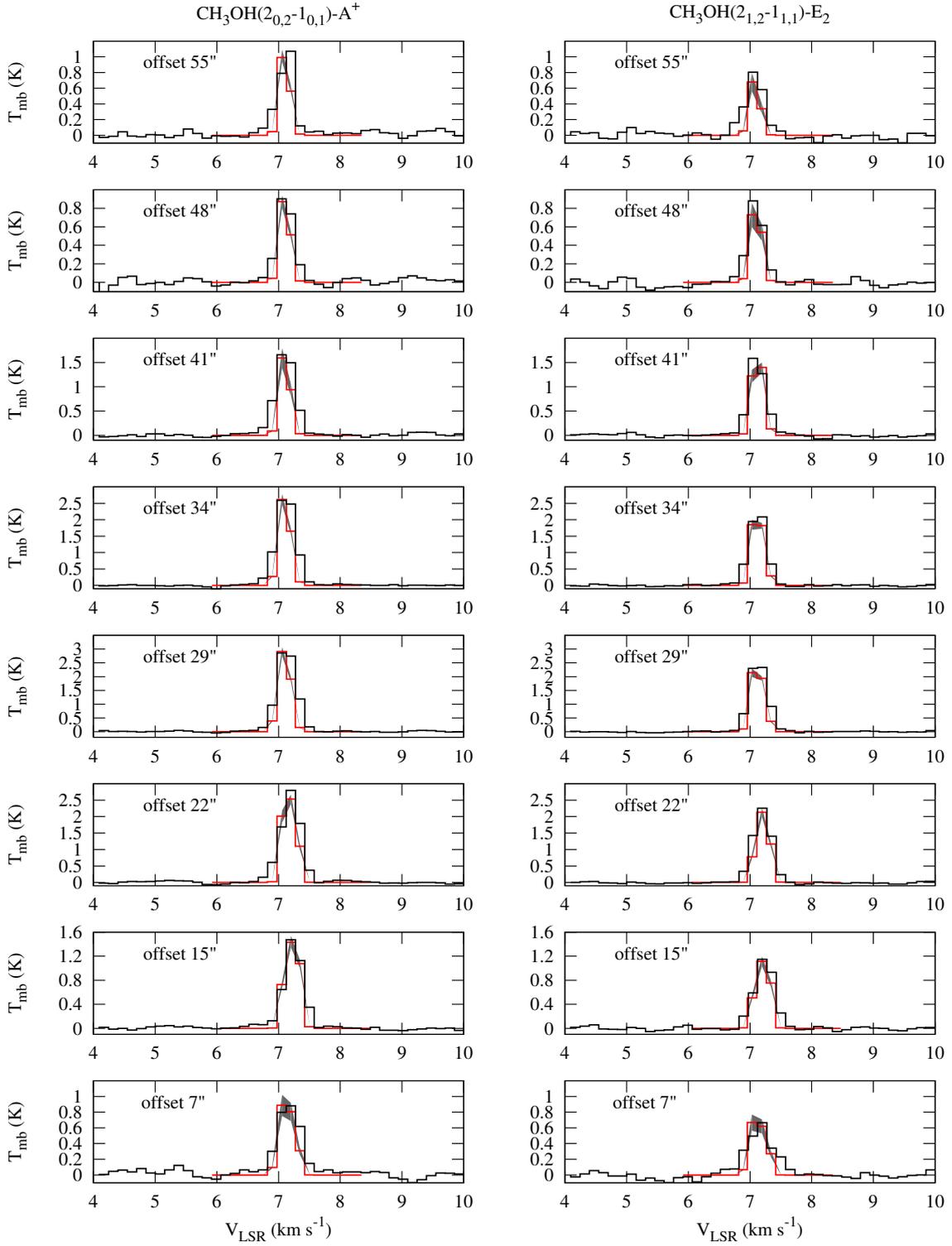

**Figure 11.** Observed and modelled spectra of the $A^+$ line (left) and $E_2$ line (right) towards all eight investigated points. The black lines show the observed spectra. The red lines show the best modelled lines. The grey strips show the lines modelled with the neighbouring knots of the grid.

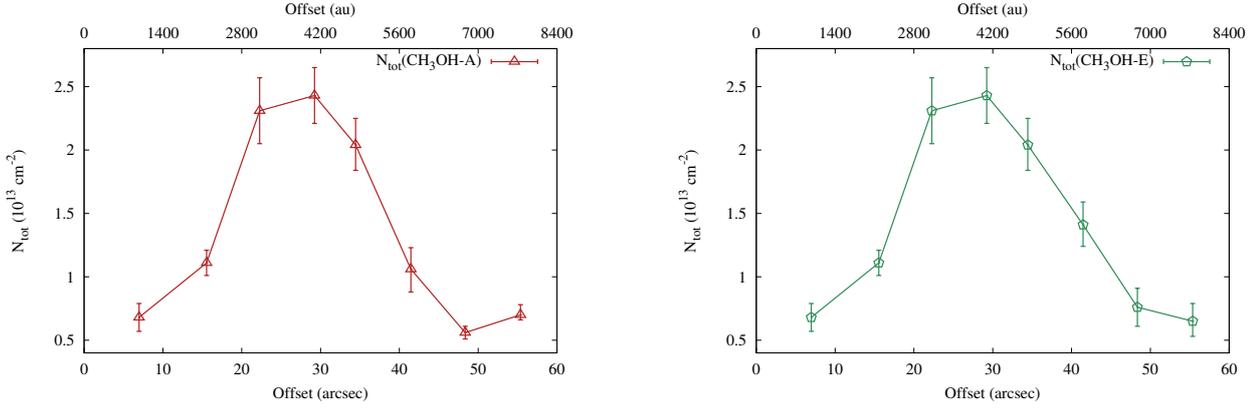

**Figure 12.** Modelled column densities of CH$_3$OH-*A* (left) and CH$_3$OH-*E* (right) towards eight selected positions across the core as a function of distance from the dust peak. The errorbars show the dispersion between the modelled column densities in the neighbouring knots of the grid.

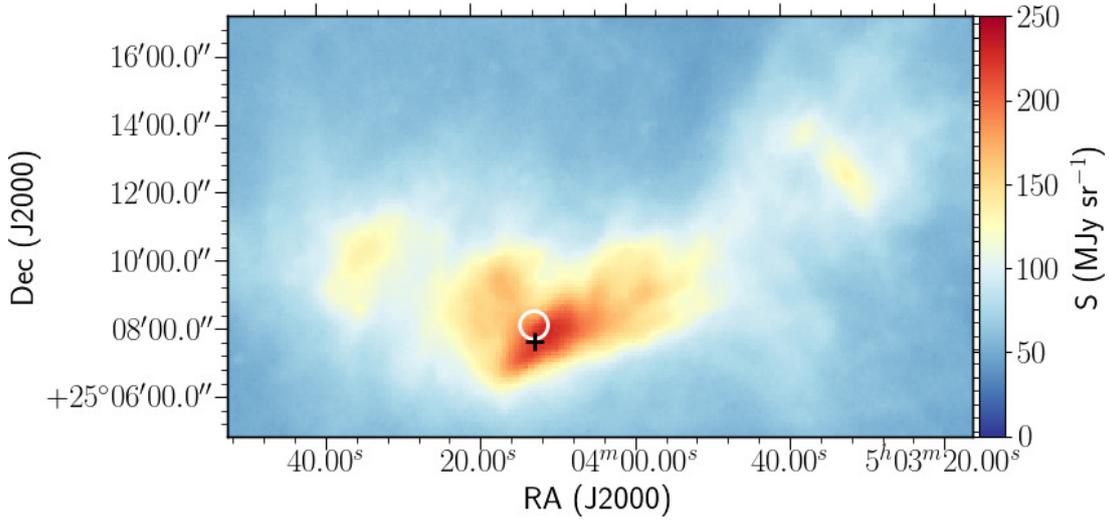

**Figure 13.** The 250 $\mu$m dust continuum emission *Herschel*/SPIRE map towards the L1544 region (André et al. 2010). The white circle shows the NOEMA primary beam centred at the methanol peak. The black cross shows the 1.3 mm dust emission peak (Ward-Thompson et al. 1999).

Fig. 9). The low accuracies of the LTE- column densities allow large variations within the errors. However, the LTE-column densities of the $A^+$ line shown in Fig. 9 present the same column densities range $(0.7$–$3.0) \times 10^{13}$ cm$^{-2}$ as the non-LTE profile (see Fig. 12). The results of non-LTE modelling give lower column density close to the dust peak (7″ away from the dust peak) compared to the results of Bizzocchi et al. (2014) and Vastel et al. (2014): $N_{tot}(\mathrm{CH_3OH}) = (2.7 \pm 0.6) \times 10^{13}$ cm$^{-2}$ and 2.6–3.8$\times 10^{13}$ cm$^{-2}$, respectively, while we derive $0.68 \times 10^{13}$ cm$^{-2}$. This difference is likely due to the larger IRAM 30 m beam of 25.5″ used in both Bizzocchi et al. (2014) and Vastel et al. (2014) works, which dilutes the area where methanol is depleted.

## 4. DISCUSSION

The single-dish observations of methanol towards L1544 (Bizzocchi et al. 2014) revealed an asymmetric ring-like structure of emission with a peak on the north-east side of the core. NOEMA has looked in detail at the CH$_3$OH peak and found morphological substructure within the NOEMA primary beam, which cannot be resolved with the IRAM 30 m telescope. The interferometric observations have revealed a complex velocity structure. A clear velocity gradient with magnitude $\simeq 7$ km s$^{-1}$ pc$^{-1}$ is found and the velocity dispersion increases towards the south-east of the NOEMA image, reaching values larger by 0.05–0.1 km s$^{-1}$ when compared to those measured with the single-dish. The higher velocity dispersion gas is concentrated towards the north-east edge of the pre-stellar core and may be produced by cloud material slowly accreting onto the core. Another possibility could be a transonic shock produced by the collision of the two almost perpendicular filaments seen in the large scale *Herschel* dust continuum map (see Fig. 13), with the L1544 methanol peak found at the intersection of the two. The currently available methanol data, however, are not sufficient to identify conclusively the presence of a shock at this location, for example with the finding of local gas temperature increase, as the temperature of the gas traced by CH$_3$OH cannot be measured. In fact, the observed transitions have low energies (see Tab. 1) and thus are not sensitive to high temperatures, which are expected even in slow shocks (see e.g. Pon et al. 2014). In the future, we plan to search for higher excitation lines of CH$_3$OH and CO to investigate this further.

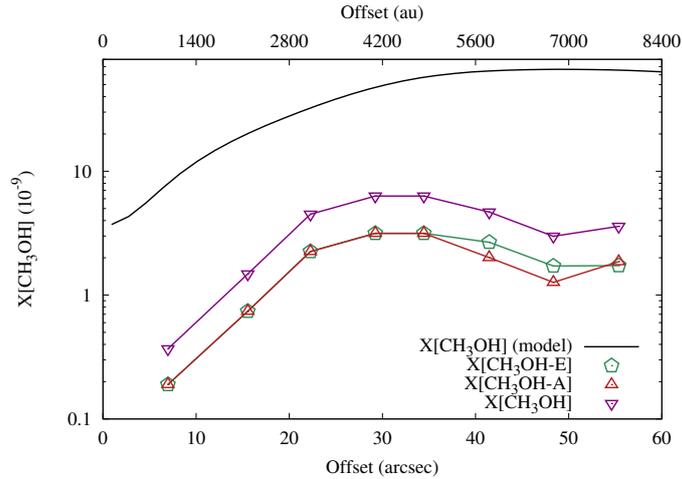

**Figure 14.** The abundances of CH$_3$OH from non-LTE modelling with MOLLIE (A shown with red, E with green, and the total abundance, the sum of A-methanol and E-methanol abundances shown with purple) and the methanol abundances predicted by Vasyunin et al. (2017) (black line).

The methanol column densities calculated with an assumption of LTE and determined with non-LTE modelling agree within the errors, but the column density close to the dust peak is two – three times smaller than that defined with the single-dish observations by Bizzocchi et al. (2014) and Vastel et al. (2014). This can be explained by large 30 m beam which partly dilutes the zone of depleted methanol. The rotational temperature calculations show that methanol is close to LTE only near the dust peak where it is significantly depleted. We compare the non-LTE molecular abundance profile to the modelled methanol abundance profile, predicted with the MONACO code (Vasyunin et al. 2017), convolved to 5″ beam, close to the size of NOEMA beam, (see Fig. 14) and find one order of magnitude lower abundances than those predicted by the model. The shape of the derived abundance profile is consistent with the model predictions, although the model suggests a slight decrease in the abundance with the radius after the maximum at ∼7000 au while the derived abundances start to decrease after the methanol peak at the radius of ∼4000 au. This is considered as fair agreement given that the modelled abundance accuracy is one order of magnitude (Vasyunin et al. 2004, 2008), however also suggests that the model over-produces the CH$_3$OH possibly by overestimating the efficiency of reactive desorption, based on the laboratory results of Minissale et al. (2016a), or by overestimating the efficiency of atomic hydrogen tunnelling through activation barriers of surface reactions taken from the work of Hasegawa et al. (1992). Another factor which can lead to the overproduction of methanol is that the static model does not allow dust temperature to exceed 10 K which prevents formation of some major ice species such as CO$_2$, where a significant fraction of carbon and oxygen may be locked. That is inclusion of physical evolution of L1544 in the model can lead to efficient formation of CO$_2$ ice, and reduced abundances of other C- and O-bearing species, including CH$_3$OH. It is also important to stress that the chemical model, as well as the radiative transfer analysis, assume spherical symmetry, thus neglect the elongated structure and differential illumination described in Spezzano et al. (2016).

## 5. SUMMARY

This paper presents high spatial resolution (∼700 au) observations of the methanol emission peak towards the prototypical pre-stellar core L1544 (revealed by Bizzocchi et al. 2014). The asymmetry of methanol emission around the dust peak of the pre-stellar core is likely caused by an irregular distribution of the core material and a lack of UV radiation at the methanol peak, as was suggested by Spezzano et al. (2016). NOEMA shows that the methanol peak has a smooth morphology, but reveals a complex velocity field. The increase in velocity dispersion towards the north-east edge of the pre-stellar core, where the local velocity gradients also present sharp changes in magnitude and direction, suggests that slow shocks are present. These slow shocks could be produced by either the accretion of cloud material onto the core or by a collision of the two filamentary structures seen in *Herschel*/SPIRE images. The NOEMA observations, coupled with a non-LTE radiative transfer analysis which takes into account the physical structure of the pre-stellar core and surroundings, also helped to unveil the methanol distribution along the line of sight; we deduced a (factor of 2–3) higher depletion of $CH_3OH$ close to the dust peak, when compared with results from single-dish observations. Comparison of the deduced $CH_3OH$ column densities with a chemical model applied to the L1544 physical structure suggests that the model is over-predicting the $CH_3OH$ abundance, probably because of a too efficient reactive desorption mechanism (which releases $CH_3OH$ molecules in the gas phase upon formation on dust grain surfaces), atomic hydrogen tunnelling or lack of physical evolution of the core in the model.

This work is part of the NOEMA large program SOLIS (Seeds of Life in Space), aimed at studying the formation of complex organic molecules at all stages of star formation (Ceccarelli et al. 2017).

The authors thank the anonymous referee for valuable comments which helped to improve the manuscript. The authors acknowledge the financial support of the European Research Council (ERC; project PALs 320620); Andy Pon acknowledges that partial salary support was provided by a CITA National Fellowship. I.J.-S. and D. Q. acknowledge the financial support received from the STFC through an Ernest Rutherford Fellowship and Grant (proposals number ST/L004801 and ST/M004139). Cecilia Ceccarelli acknowledges the financial support of the ERC (project DOC 741002).

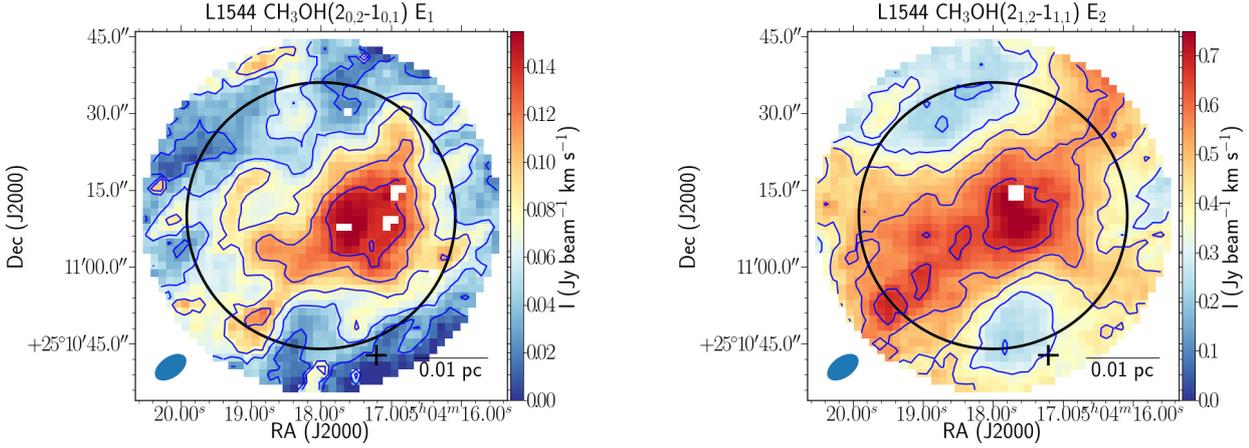

**Figure A.1.** Integrated intensities of the methanol lines (NOEMA+30 m) for the $E_1$ (left) and $E_2$ (right) lines. The blue contours represent integrated intensity, and start at 0.027 Jy beam$^{-1}$ km s$^{-1}$ with a step of 0.027 Jy beam$^{-1}$ km s$^{-1}$ for the $E_1$ line (left), and at 0.108 Jy beam$^{-1}$ km s$^{-1}$ with a step of 0.108 Jy beam$^{-1}$ km s$^{-1}$ for the $E_2$ line (right). $3\sigma_I$=0.005 Jy beam$^{-1}$ km s$^{-1}$. The circle shows the primary beam of NOEMA. The cross shows the dust emission peak (Ward-Thompson et al. 1999). The synthesized beam of NOEMA is shown in the bottom left corner. The white pixels are those masked because of low quality spectra.

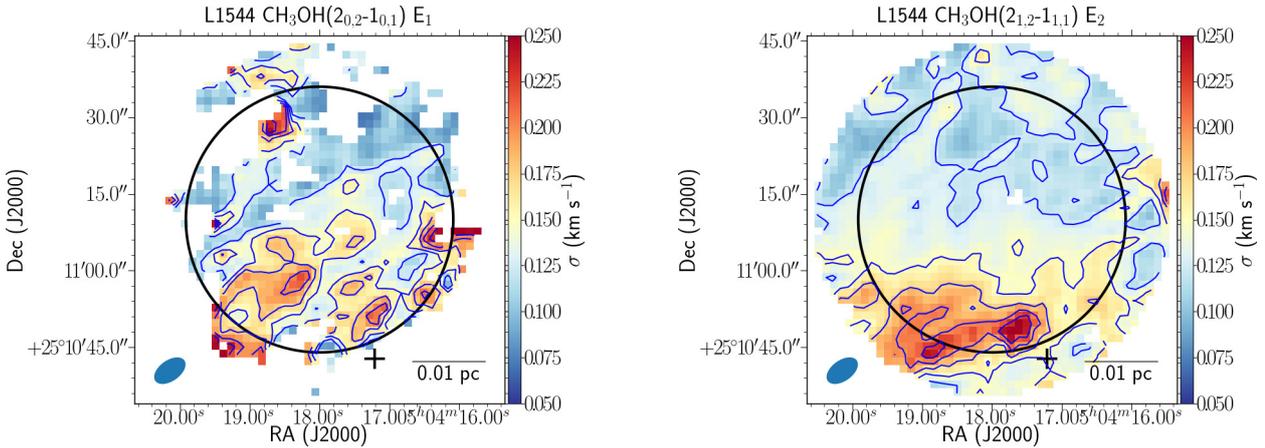

**Figure A.2.** Velocity dispersions of the methanol lines for the $E_1$ (left) and $E_2$ (right) lines. The blue contours show velocity dispersions of 0.125, 0.150, 0.175, 0.200, and 0.225 km s$^{-1}$. The circle shows the primary beam. The cross shows the 1.3 mm dust emission peak (Ward-Thompson et al. 1999). The synthesized beam is shown in the bottom left corner. The white pixels are those masked because of low quality spectra.

# APPENDIX

## A. THE ADDITIONAL FIGURES

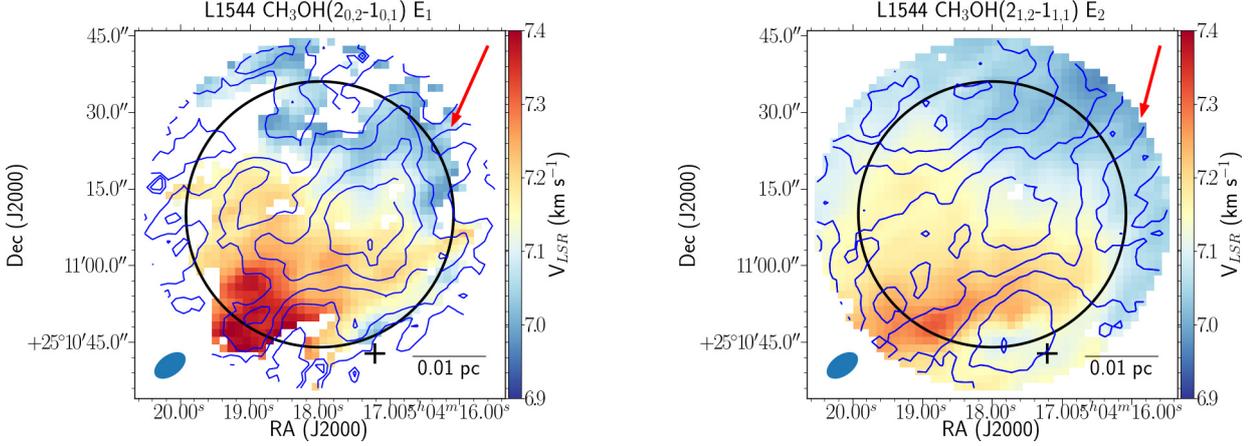

**Figure A.3.** Centroid velocities of the methanol lines for the $E_1$ (left) and $E_2$ (right) lines. The blue contours represent integrated intensity, and start at 0.027 Jy beam$^{-1}$ km s$^{-1}$ with a step of 0.027 Jy beam$^{-1}$ km s$^{-1}$ for the $E_1$ line (left), and at 0.108 Jy beam$^{-1}$ km s$^{-1}$ with a step of 0.108 Jy beam$^{-1}$ km s$^{-1}$ for the $E_2$ line (right). $3\sigma_I$=0.005 Jy beam$^{-1}$ km s$^{-1}$. The red arrows show the total velocity gradients (8.77$\pm$0.04 and 7.24$\pm$0.01 km s$^{-1}$ pc$^{-1}$ for the $E_1$ and $E_2$ lines). The circle shows the primary beam. The cross shows the 1.3 mm dust emission peak (Ward-Thompson et al. 1999). The synthesized beam is shown in the bottom left corner. The white pixels are those masked because of low quality spectra.

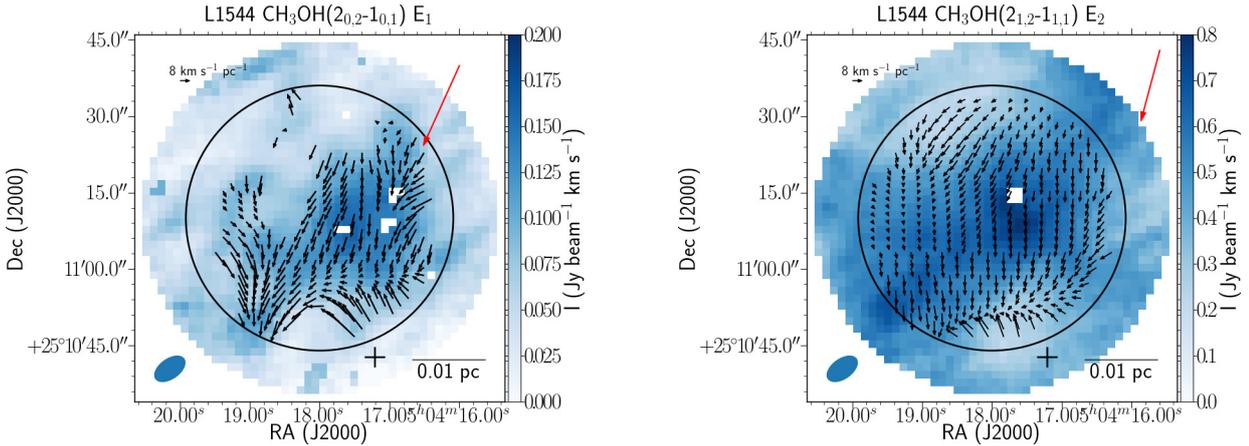

**Figure A.4.** Local velocity gradients of the methanol lines for the $E_1$ (left) and $E_2$ (right) lines. The color scale shows integrated intensity. The black arrows indicate the local velocity gradients. The red arrows show the total velocity gradients (the scale of the total gradients are eight times larger than the scale of the local gradients). The circle shows the primary beam. The cross denotes the 1.3 mm dust emission peak (Ward-Thompson et al. 1999). The synthesized beam is plotted in the bottom left corner. The white pixels are those masked because of low quality spectra.